\documentclass[11pt,a4paper]{article}

\usepackage[T1]{fontenc}
\usepackage[utf8]{inputenc}
\usepackage{lmodern}
\usepackage{amsmath,amssymb}
\usepackage{geometry}
\usepackage{xcolor}
\usepackage{microtype}
\usepackage{cite}
\usepackage[hypertexnames=false]{hyperref}

\hypersetup{
	colorlinks=true,
	linkcolor=blue,
	citecolor=blue,
	urlcolor=blue,
	pdftitle={A Darboux-Theorem-Based Derivation of Geometric Structures in Various Non-Abelian Gauge Theories},
	pdfauthor={Stefan-Sabin Manolescu and Eugen-Mihaita Cioroianu}
}

\newcommand{\keywords}[1]{%
	\par\medskip
	\noindent\textbf{Keywords:} #1
}

\numberwithin{equation}{section}

\begin{document}

\title{A Darboux-Theorem-Based Derivation of Geometric Structures in Various Non-Abelian Gauge Theories
}

\author{%
Stefan-Sabin Manolescu\\
\small Doctoral School of Sciences, University of Craiova\\
\small 13 A.I. Cuza Street, 200585 Craiova, Romania\\
\small \texttt{manolescu.stefan.y3a@student.ucv.ro}
\and
Eugen-Mihaita Cioroianu\\
\small Department of Physics, University of Craiova\\
\small 13 A.I. Cuza Street, 200585 Craiova, Romania\\
\small \texttt{cioroianu.eugen@ucv.ro}
}

\date{}

\maketitle

\begin{abstract}
This paper illustrates the straightforward application of the Faddeev--Jackiw approach to several non-Abelian gauge theories, namely, the Freedman--Townsend, Yang--Mills, and non-Abelian BF models, as well as a nonlinear theory of BF type. For each theory, the procedure determines the phase space and its symplectic structure, the coisotropic constraint submanifold, the Hamiltonian, and the reducibility properties of the constraints. By combining these results with the Dirac conjecture, one reconstructs through direct computation the corresponding Lagrangian gauge transformations and their reducibility structure. The analysis provides a unified and efficient derivation of the principal Hamiltonian and Lagrangian gauge structures of these non-Abelian theories.
\end{abstract}



\section{Motivation}	
Developed in the 1980s, the Faddeev--Jackiw approach \cite{Faddeev} provides an economical method for determining the Poisson structure on phase space, the dynamically accessible coisotropic submanifold, and the generator of time evolution associated with a given first-order Lagrangian system. These structures constitute essential ingredients of the canonical quantization framework \cite{Berezin}. Subsequently, one of the originators of the method provided an elegant geometric justification of the prescription based on the Darboux theorem \cite{Libermann}, which guarantees the local existence of canonical coordinates for a symplectic structure and, correspondingly, a canonical local form of the associated symplectic potential \cite{Jackiw}. In the ensuing years, the Faddeev--Jackiw approach inspired several related analyses of general first-order systems \cite{Govaerts,Barcelos1,Barcelos2,Long}. 

Although its underlying strategy is relatively simple, the Faddeev--Jackiw approach has, to the best of our knowledge, been applied primarily to toy models with finitely many degrees of freedom and to linear field theories \cite{Cioroianu}. This observation provides the main motivation for investigating the Hamiltonian formulations of several non-Abelian theories within the Faddeev--Jackiw framework. The models considered here occupy a prominent place in modern gauge theory and comprise (i) the non-Abelian Freedman--Townsend theory, a paradigmatic model of antisymmetric tensor gauge fields that is closely related, through duality, to nonlinear principal chiral sigma models \cite{FT}; (ii) Yang--Mills theory, which provides the fundamental gauge-theoretic framework underlying both the electroweak and strong interactions \cite{YM,Weinberg,GrossWilczek,Politzer}; (iii) non-Abelian BF theory, which belongs to a distinguished class of metric-independent topological gauge theories and provides a natural framework for studying moduli spaces of flat connections, topological quantum field theories, and BF-type formulations of classical and quantum gravity \cite{BT,Horowitz,BaezBF,Plebanski}; and (iv) a nonlinear topological model of BF type \cite{TOP}, belonging to a broader family of nonlinear topological gauge theories that includes Poisson sigma models \cite{Ikeda,Schaller,CattaneoFelder}, Courant sigma models \cite{Roytenberg}, and various higher gauge theories, and that provides a natural setting for the Batalin--Vilkovisky formalism through the AKSZ construction \cite{AKSZK}.

To preserve generality, these theories are formulated on $D$-dimensional Minkowski spacetime, $\mathbb{R}^{1,D-1}$, endowed with the mostly-minus signature. This general-dimensional setting reveals higher-stage reducible gauge structures in all the models considered, with the sole exception of Yang--Mills theory.

The paper is organized into five sections. Section 2 provides a concise review of the Faddeev--Jackiw procedure, presented, for simplicity, in the setting of a generic bosonic first-order system with finitely many degrees of freedom. In Section 3, this general framework is applied to three non-Abelian gauge theories: the Freedman--Townsend theory \cite{FT}, Yang--Mills theory \cite{YM}, and non-Abelian BF theory \cite{BT}. The first two models are based on a (compact) semisimple Lie algebra, whereas the third is formulated in terms of a general non-Abelian Lie algebra and its dual vector space. By virtue of their Lagrangian formulations, the Freedman--Townsend and non-Abelian BF models are already first order in the field derivatives and are therefore directly amenable to the Faddeev--Jackiw procedure. By contrast, the Yang--Mills Lagrangian is quadratic in the field velocities; accordingly, its Faddeev--Jackiw analysis requires a preliminary first-order reformulation achieved through the introduction of auxiliary fields. For the Freedman--Townsend model, the procedure yields an Abelian Hamiltonian gauge algebra that is on-shell $(D-3)$-stage reducible. In the Yang--Mills case, it produces an irreducible non-Abelian Hamiltonian gauge algebra, whereas, for the non-Abelian BF model, it leads to a non-Abelian Hamiltonian gauge algebra that is on-shell $(D-3)$-stage reducible. Section 4 is devoted to the Faddeev--Jackiw analysis of a nonlinear topological model of BF type, whose formulation is governed from the outset by a Poisson structure on the target space parametrized by the scalar fields. The resulting Hamiltonian gauge algebra is genuinely nonlinear, closes only on-shell, and is itself on-shell $(D-2)$-stage reducible. For each of the four models, the analysis culminates in the reconstruction of the corresponding Lagrangian gauge structure by invoking the Dirac conjecture, according to which first-class constraints generate gauge transformations. Finally, Section 5 summarizes the principal results and presents the concluding remarks. 
\section{Darboux's theorem at work: The Faddeev--Jackiw--Dirac approach}\label{GenTh}
The Faddeev--Jackiw analysis \cite{Faddeev} of a first-order system, taken to be bosonic for simplicity and described by the action
\begin{equation}\label{FJ1}
	S[z]=\int{\rm d}t\left(\alpha_A\dot{z}^A-V(z)\right),
\end{equation}
relies \cite{Jackiw} on the presymplectic Darboux theorem \cite{Libermann}, applied to the exterior derivative of the kinetic $1$-form
\begin{equation}\label{FJ2}
	{\rm d}\alpha,\quad\alpha:=\alpha_A{\rm d}z^A.
\end{equation}
The theorem applies whenever the closed $2$-form ${\rm d}\alpha$ has constant rank, say
\begin{equation}\label{FJ3}
	{\rm rank}({\rm d}\alpha)=2k.
\end{equation}

Under this assumption, the presymplectic Darboux theorem guarantees the existence of a local parametrization
\[
(z^A)\longleftrightarrow(q^i,p_i,Q^\Delta)
\]
in which  ${\rm d}\alpha$ assumes its canonical form on the symplectic directions, while the variables $Q^\Delta$ parametrize its kernel. Up to the addition of a total time derivative, the action \eqref{FJ1} can therefore be written as
\begin{equation}\label{FJ4}
S[q,p,Q]=\int{\rm d}t\left(p_i\dot{q}^i-\mathcal{V}(q,p,Q)\right).
\end{equation}
The null variables may then be separated as 
\[
(Q^\Delta)=(Q^\mu,u^a)
\]
where the $Q^\mu$ are auxiliary variables, whereas the $u^a$ remain undetermined and eventually play the role of Lagrange multipliers. The auxiliary variables are eliminated through their algebraic equations of motion,
\begin{equation}\label{FJ5}
	\frac{\partial\mathcal{V}(q,p,Q)}{\partial Q^\mu}=0\Longleftrightarrow Q^\mu=\Psi^\mu(q,p,u).
\end{equation} 
Upon substituting the solutions \eqref{FJ5} into \eqref{FJ4}, and after performing any further necessary redefinitions of the null variables, the original action can be brought to the form
\begin{equation}\label{FJ6}
S[q,p,u]=\int{\rm d}t\left(p_i\dot{q}^i-h(q,p)-u^aG_a(q,p)\right).
\end{equation}

If necessary, one subsequently restricts the system to the submanifold defined by those constraints that can be solved and eliminates the corresponding variables. The remaining constraints,
\begin{equation}\label{FJ7}
	G_a\approx0,
\end{equation}
can then be taken to be first class,
\[
\left[G_a,G_b\right]\approx0,
\]
while the time-evolution generator, $h(q,p)$, is first-class with respect to the same set of constraints \eqref{FJ7},
\[
\left[h,G_a\right]\approx0.
\]

At this stage, the analysis can be completed either by solving the remaining constraints \cite{Faddeev} or by invoking Dirac's conjecture \cite{Marc} to derive the Hamiltonian gauge algebra associated with \eqref{FJ6}. In field-theoretic settings such as the one considered here, where locality is a fundamental requirement, the latter option is preferable.
\section{Non-Abelian theories}
The present section is devoted to the Faddeev--Jackiw analysis of several standard non-Abelian gauge theories. These models involve Lie-algebra-valued one-forms associated with semisimple Lie algebras and possess gauge degrees of freedom. The general procedure developed in Section \ref{GenTh} allows one to determine, in a straightforward manner, the phase space, its symplectic structure, the auxiliary variables, the coisotropic constraint submanifold, and the generator of time evolution, namely, the first-class Hamiltonian, for each of the theories under consideration. Furthermore, when supplemented by the Dirac conjecture, the Hamiltonian results furnished by the Faddeev--Jackiw approach allow one to reconstruct the complete Lagrangian gauge structure of each original model.
\subsection{Non-Abelian Freedman-Townsend model} 
The first model, namely the non-Abelian Freedman--Townsend theory \cite{FT}, provides a consistent non-Abelian generalization of Abelian two-form gauge theories and is dual to a class of nonlinear sigma models.
At the Lagrangian level, its evolution comes from the variational principle associated with the Lagrangian
\begin{equation}\label{FT1}
	L^{\rm FT}(t):=\tfrac{1}{2}\int{\rm d}^{D-1}{\bf{x}}\left(B^{\mu\nu}_aF^a_{\mu\nu}+A^a_\mu A^\mu_a\right),
\end{equation}
where
\begin{equation}\label{FSnonA}
F^a_{\mu\nu}:=\partial_{[\mu}A^a_{\nu]}+f^a_{bc}A^b_\mu A^c_\nu
\end{equation}
denotes the non-Abelian field-strength tensor associated with the gauge field $A_\mu^a$, with $f^a_{bc}$ being the structure constants of the underlying $N$-dimensional semisimple Lie algebra $\mathfrak{g}$. The quantities $B^{\mu\nu}_a=-B^{\nu\mu}_a$ represent the components of a $\mathfrak{g}$-valued two-form field. It is worth recalling that the semisimplicity of $\mathfrak{g}$ is equivalent to the non-degeneracy of its Killing--Cartan form, which, in turn, induces a canonical identification between $\mathfrak{g}$ and its dual space $\mathfrak{g}^\ast$.

According to the general procedure, upon decomposing the spacetime indices into their temporal and spatial components and performing appropriate partial integrations, the functional \eqref{FT1} takes the form
\begin{equation}\label{FT2}
	L^{\rm FT}(t):=\int{\rm d}^{D-1}{\bf{x}}\left(B^{0j}_a\dot{A}^a_j+A^a_0(D_j)_a^{\hphantom{a}b}B^{0j}_b+\tfrac{1}{2}A^a_0A^0_a+\tfrac{1}{2}B^{jk}_aF^a_{jk}+\tfrac{1}{2}A^a_j A^j_a\right),
\end{equation}
where the standard notations
\begin{equation}\label{FT3}
	(D_j)_a^{\hphantom{a}b}:=\delta^b_a\partial_j+f^b_{ac}A^c_j,\quad (D_j)^a_{\hphantom{a}b}:=\delta_b^a\partial_j-f^a_{bc}A^c_j
\end{equation}
have been employed for the covariant derivatives.

Inspection of \eqref{FT2} immediately reveals that the fields $A^a_0$ are purely auxiliary, since their Euler--Lagrange equations take the form
\begin{equation}\label{FT4}
\frac{\delta L^{\rm FT}}{\delta A^a_0}\equiv(D_j)_a^{\hphantom{a}b}B^{0j}_b+A^0_a=0\Longleftrightarrow A^0_a=-(D_j)_a^{\hphantom{a}b}B^{0j}_b.
\end{equation}
Taking advantage of the auxiliary character of the temporal components of the gauge field, one may eliminate them by substituting their equations of motion \eqref{FT4} back into the Lagrangian, thereby obtaining the reduced functional
\begin{equation}\label{FT5}
	\bar{L}^{\rm FT}(t):=\int{\rm d}^{D-1}{\bf{x}}\left(B^{0j}_a\dot{A}^a_j-\tfrac{1}{2}(D^k)^a_{\hphantom{a}c}B^c_{0k}(D_j)_a^{\hphantom{a}b}B^{0j}_b+\tfrac{1}{2}A^a_j A^j_a+\tfrac{1}{2}B^{jk}_aF^a_{jk}\right).
\end{equation}

At this point, using the prescriptions of Faddeev--Jackiw approach applied on the Lagrangian functional \eqref{FT5}, one derives the following results. First, Freedman-Townsend model exhibits a phase space parametrized by the pair $(B^{0k}_a,A^a_k)$, and is equipped with a non-degenerate Poisson $2$-vector whose only non-vanishing fundamental brackets are
\begin{equation}\label{FT6}
	\left[B^{0k}_a,A^b_l\right]=-\delta^b_a\delta^k_l.
\end{equation}
Second, the fields $B^{jk}_a$ act as Lagrange multipliers for the functions $\tfrac{1}{2}F^a_{jk}$, implying that the dynamics is restricted to the constraint surface
\begin{equation}\label{FT7}
	G^a_{jk}:=\tfrac{1}{2}F^a_{jk}\approx 0.
\end{equation}
The explicit form of the field-strength components further shows that these constraints constitute an Abelian first-class set,
\[
\left[G^a_{jk},G^b_{lm}\right]=0.
\]
Moreover, the same observation implies that the constraints \eqref{FT7} are on-shell $(D-3)$-reducible. The first-order reducibility relations 
\begin{equation}\label{FT8}
	\left(Z^a_{jkl}\right)_b^{mn}G^b_{mn}=0,\quad\left(Z^a_{jkl}\right)_b^{mn}:=\tfrac{1}{3!}(D_{[j})^a_{\hphantom{a}b}\delta^m_k\delta^n_{l]},
\end{equation}
whereas the remaining reducibility relations take the form
\begin{equation}\label{FT9}
	\left(Z^a_{k_1\cdots k_{p}}\right)_b^{l_1\cdots l_{p-1}}\left(Z^b_{l_1\cdots l_{p-1}}\right)_c^{j_1\cdots j_{p-2}}=-\tfrac{2}{p!}f^a_{cd}G^d_{[k_1k_2}\delta^{j_1}_{k_3}\cdots\delta^{j_{p-2}}_{k_p]},
\end{equation}
with the $p$-order reducibility functions defined by
\begin{equation}\label{FT9r}
\left(Z^a_{k_1\cdots k_{p}}\right)_b^{l_1\cdots l_{p-1}}:=\tfrac{1}{p!}(D_{[k_1})^a_{\hphantom{a}b}\delta^{l_1}_{k_2}\cdots\delta^{l_{p-1}}_{k_p]},\quad p=3,4,\cdots,(D-1).
\end{equation}

Third, the canonical Hamiltonian extracted from \eqref{FT5}
\begin{equation}\label{FT10}
\bar{H}^{\rm FT}(t):=\tfrac{1}{2}\int{\rm d}^{D-1}{\bf{x}}\left((D^k)^a_{\hphantom{a}c}B^c_{0k}(D_j)_a^{\hphantom{a}b}B^{0j}_b-A^a_j A^j_a\right)
\end{equation}
coincides with the generator of time evolution, as it is a first-class function with respect to the constraint set \eqref{FT7},
\begin{equation}\label{FT11}
	\left[G^a_{jk},H\right]=f^a_{bc}G^b_{jk}(D^l)^c_{\hphantom{c}d}B^d_{0l}.
\end{equation}

Fourth, a simple counting argument based on the dimensions of the phase space and the first-class constraint surface reveals that the model under consideration propagates $N$ physical degrees of freedom per spatial point. Remarkably, this number is {\em independent} of the dimension of the underlying Minkowski spacetime.

The standard Dirac prescription, according to which first-class constraints act as generators of gauge transformations \cite{Marc}, directly yields the generating set of gauge transformations associated with the action functional
\begin{equation}\label{FT12}
\bar{S}^{\rm FT}:=\int{\rm d}t \bar{L}^{\rm FT}(t)
\end{equation}
namely
\begin{equation}\label{FT13}
	\bar{\delta}_\epsilon A^a_j=0,\quad\bar{\delta}_\epsilon B^{0j}_a=(D_k)_a^{\hphantom{a}b}\epsilon^{kj}_b,
\end{equation}
together with
\begin{equation}\label{FT14}
\delta_\epsilon \bar{B}^{jk}_a=-\dot{\epsilon}^{jk}_a-f^b_{ac}(D^l)^c_{\hphantom{c}d}B_{0l}^d\epsilon^{jk}_b+(D_l)_a^{\hphantom{a}b}\epsilon^{ljk}_b
\end{equation}
where the latter transformation is required by the invariance of the action \eqref{FT12}, in conjunction with the first-stage reducibility relations \eqref{FT8} satisfied by the first-class constraints \eqref{FT7}.

Finally, returning to the covariant Lagrangian action associated with \eqref{FT1}, or, equivalently, with \eqref{FT2}, 
\begin{equation}\label{FT15}
	S^{\rm FT}:=\int{\rm d}t L^{\rm FT}(t)
\end{equation}
one can infer from the local invariances \eqref{FT13}--\eqref{FT14} the covariant generating set of gauge transformations
\begin{equation}\label{FT16}
	\delta_\epsilon A^a_\mu=0,\quad \delta_\epsilon B_a^{\mu\nu}=(D_\lambda)_a^{\hphantom{a}b}\epsilon^{\lambda\mu\nu}_b
\end{equation}
subject to the identification of gauge parameters
\[
\epsilon_a^{0jk}:=-\epsilon^{jk}_a.
\]
Furthermore, the generating set of gauge transformations \eqref{FT16} defines an Abelian gauge algebra,
\[
\left[\delta_\epsilon,\delta_\eta\right]A^a_\mu=0=\left[\delta_\epsilon,\delta_\eta\right]B^{\mu\nu}_a
\]
and is on-shell $(D-3)$-stage reducible, with the first-stage reducibility relations
\begin{equation}\label{FT17}
(Z^{\mu\nu}_a)^b_{\alpha\beta\gamma}(Z^{\alpha\beta\gamma}_b)^c_{\lambda\rho\sigma\pi}=-\tfrac{2}{4!}f^c_{ab}\frac{\delta S^{\rm FT}}{\delta B_b^{[\lambda\rho}}\delta^\mu_\sigma\delta^\nu_{\pi]},
\end{equation}
where
\begin{equation}\label{FT18}
(Z^{\mu\nu}_a)^b_{\alpha\beta\gamma}:=\tfrac{1}{3!}(D_{[\alpha})_a^{\hphantom{a}b}\delta^\mu_\beta\delta^\nu_{\gamma]},\quad (Z^{\mu\nu\rho}_a)^b_{\alpha\beta\gamma\delta}:=\tfrac{1}{4!}(D_{[\alpha})_a^{\hphantom{a}b}\delta^\mu_\beta\delta^\nu_{\gamma}\delta^\rho_{\delta]}
\end{equation}
denote the gauge generators and the first-stage reducibility functions, respectively.
The higher-stage reducibility relations take the explicit form
\begin{equation}\label{FT19}
(Z^{\mu_1\cdots\mu_{p-2}}_a)^b_{\nu_1\cdots\nu_{p-1}}(Z^{\nu_1\cdots\nu_{p-1}}_b)^c_{\rho_1\cdots\rho_p}=-\tfrac{2}{p!}f^c_{ab}\frac{\delta S^{\rm FT}}{\delta B_b^{[\rho_1\rho_2}}\delta^{\mu_1}_{\rho_3}\cdots\delta^{\mu_{p-2}}_{\rho_p]},\quad p=4,\cdots,D
\end{equation}
where the corresponding higher-stage reducibility functions are given by
\begin{equation}\label{FT20}
(Z^{\mu_1\cdots\mu_{p-1}}_a)^b_{\nu_1\cdots\nu_{p}}:=\tfrac{1}{p!}(D_{[\nu_1})_a^{\hphantom{a}b}\delta^{\mu_1}_{\nu_2}\cdots\delta^{\mu_{p-1}}_{\nu_p]},\quad p=3,\cdots,(D-1).
\end{equation}
\subsection{Yang-Mills theory} 

The Yang-Mills model \cite{YM} assumes a dynamics generated, via the variational principle, by the Lagrangian
\begin{equation}\label{YM1}
	L^{\rm YM}(t):=\int{\rm d}^{D-1}{\bf x}-\tfrac{1}{4}F^a_{\mu\nu}F_a^{\mu\nu},
\end{equation}
with the field-strength given previously \eqref{FSnonA}. 

To implement the Faddeev--Jackiw procedure, one first rewrites \eqref{YM1} by performing a space--time decomposition, obtaining
\begin{equation}\label{YM2}
	L^{\rm YM}(t):=\int{\rm d}^{D-1}{\bf x}\left(-\tfrac{1}{2}F^a_{0k}F_a^{0k}-\tfrac{1}{4}F^a_{jk}F_a^{jk}\right).
\end{equation}
The temporal derivatives are then linearized through the introduction of suitable auxiliary fields, which leads to the equivalent first-order Lagrangian
\begin{equation}\label{YM3}
\bar{L}^{\rm YM}(t):=\int{\rm d}^{D-1}{\bf x}\left(B^k_aF^a_{0k}+\tfrac{1}{2}B^a_kB^k_a-\tfrac{1}{4}F^a_{jk}F_a^{jk}\right).
\end{equation} 
Indeed, the Euler--Lagrange equations corresponding to the auxiliary fields read
\begin{equation}\label{YM4}
	\frac{\delta \bar{L}^{\rm YM}}{\delta B^k_a}\equiv B^a_k+F^a_{0k}=0\Longleftrightarrow B^a_k=-F^a_{0k},
\end{equation}
thereby establishing the equivalence between the first-order formulation \eqref{YM3} and the original Lagrangian \eqref{YM2}.

At this stage, a straightforward integration by parts allows one to rewrite \eqref{YM3} in the form
\begin{equation}\label{YM5}
	\bar{L}^{\rm YM}(t):=\int{\rm d}^{D-1}{\bf x}\left(B^k_a\dot{A}^a_k+\tfrac{1}{2}B^a_kB^k_a-\tfrac{1}{4}F^a_{jk}F_a^{jk}+A^a_0(D_k)_a^{\hphantom{a}b}B^k_b\right),
\end{equation} 
which is now in a form suitable for the application of the Faddeev--Jackiw procedure. It should be noted that the covariant derivatives are defined as in \eqref{FT3}.

First, the model possesses a phase space parametrized, at each spatial point, by the pair of fields $(B^k_a,A^a_k)$ and is endowed with a non-degenerate Poisson bivector whose only non-vanishing fundamental Poisson brackets, expressed in DeWitt condensed notation, are
\begin{equation}\label{YM6}
	\left[B^k_a,A^b_l\right]=-\delta^b_a\delta^k_l.
\end{equation}
It should be noted that the temporal components $A^a_0$ of the Lie-algebra-valued one-form $A^a_\mu$ act as Lagrange multipliers for the constraint functions introduced below.

Second, the dynamics of theory is confined on the surface
\begin{equation}\label{YM7}
	G_a:=(D_k)_a^{\hphantom{a}b}B^k_b\approx 0,
\end{equation}
while the behavior of the constraint functions with respect to the canonical structure \eqref{YM6} is described by
\begin{equation}\label{YM8}
	\left[G_a,G_b\right]=f^c_{ab}G_c.
\end{equation}
These relations establish the first-class nature of the constraints and, moreover, show that the Poisson algebra of the constraint functions realizes a linear representation of the underlying Lie algebra $\mathfrak{g}$. 

Third, the time-evolution generator,
\begin{equation}\label{YM9}
	\bar{H}(t):=\tfrac{1}{2}\int{\rm d}^{D-1}{\bf x}\left(-B^a_kB^k_a+\tfrac{1}{2}F^a_{jk}F_a^{jk}\right)
\end{equation}
enjoys
\begin{equation}\label{YM10}
	\left[G_a,H\right]=0,
\end{equation}
completing in this manner the Hamiltonian gauge algebra.

Fourth, a simple counting argument based on the dimensions of the phase space and the first-class constraint surface reveals that the model under consideration propagates $(D-2)N$ physical degrees of freedom per spatial point.

Invoking Dirac's prescription for Hamiltonian gauge transformations, a straightforward computation shows that the Lagrangian action
\begin{equation}\label{YM11}
\bar{S}^{\rm YM}:=\int{\rm d}t \bar{L}^{\rm YM}(t)
\end{equation}
is invariant under the gauge transformations
\begin{equation}\label{YM12}
\bar{\delta}_\epsilon A^a_k=(D_k)^a_{\hphantom{a}b}\epsilon^b,\quad\bar{\delta}_\epsilon B^k_a=-f^c_{ab}B^k_c\epsilon^b,\quad \bar{\delta}_\epsilon A^a_0=\dot{\epsilon}^a+f^a_{bc}A^b_0\epsilon^c.
\end{equation}

Finally, after integrating out the auxiliary fields $B^a_k$ from \eqref{YM11}, one readily identifies the generating set of gauge transformations
\begin{equation}\label{YM13}
	\delta_\epsilon A^a_\mu=(D_\mu)^a_{\hphantom{a}b}\epsilon^b
\end{equation}
associated with the Lagrangian action
\begin{equation}\label{YM14}
S^{\rm YM}:=\int{\rm d}t L^{\rm YM}(t),
\end{equation}
The resulting gauge theory is irreducible, and its gauge algebra closes off-shell, with the first-order structure functions being precisely the structure constants of the underlying Lie algebra,
\begin{equation}\label{YM15}
	\left[\delta_\epsilon,\delta_\eta\right]A^a_\mu=\delta_\xi A^a_\mu,\quad \xi^a=f^a_{bc}\epsilon^b\eta^c.
\end{equation}
\subsection{Non-Abelian BF theories}
Originally developed in \cite{BT} as a class of topological field theories involving Lie-algebra-valued two-forms, BF theories on the flat Minkowski spacetime $\mathbb{R}^{1,D-1}$, with $D>2$, are described by the Lagrangian local functional
\begin{equation}\label{BF1}
	L^{\rm BF}:=\tfrac{1}{2}\int{\rm d}^{D-1}{\bf x}B^{\mu\nu}_aF^a_{\mu\nu},
\end{equation}
with the field-strength given in \eqref{FSnonA}. In contrast to the previous models, the underlying Lie algebra $\mathfrak{g}$ need not be semisimple. Accordingly, $A^a_\mu$ is a $\mathfrak{g}$-valued one-form, whereas $B^{\mu\nu}_a$ is a $\mathfrak{g}^\ast$-valued two-form.

Since the Lagrangian is already of first order, the model is directly amenable to the standard Faddeev--Jackiw procedure. Indeed, decomposing the Lorentz indices in \eqref{BF1} into their temporal and spatial components yields 
\begin{align}\label{BF2}
	L^{\rm BF}&=\int{\rm d}^{D-1}{\bf x}\left(B^{0k}_aF^a_{0k}+\tfrac{1}{2}B^{jk}_aF^a_{jk}\right)\nonumber\\
	&=\int{\rm d}^{D-1}{\bf x}\left(B^{0k}_a\dot{A}^a_k+A^a_0(D_k)_a^{\hphantom{a}b}B^{0k}_b+\tfrac{1}{2}B^{jk}_aF^a_{jk}\right),
\end{align}
from which the results summarized below follow immediately.

First, the phase space is parametrized by the pair of fields $(A^a_k,B^{0k}_a)$, while the remaining fields, $A^a_0$ and $B^{jk}_a$, act as Lagrange multipliers enforcing the constraint equations
\begin{equation}\label{BF3}
	G_a:=(D_k)_a^{\hphantom{a}b}B_b^{0k}\approx0,\quad G^a_{jk}:= \tfrac{1}{2}F^a_{jk}\approx0.
\end{equation}

Second, the phase space is endowed with the same non-degenerate Poisson $2$-vector as that of the non-Abelian Freedman--Townsend model; in particular, the only non-vanishing fundamental Poisson brackets are given by \eqref{FT6}.

Third, unlike the previous models, the non-Abelian BF theory possesses a trivial Hamiltonian dynamics, as its canonical Hamiltonian vanishes identically,
\begin{equation}\label{BF4}
	H^{\rm BF}(t)=0.
\end{equation}

By direct computation, it results that the functions \eqref{BF3} behave with respect to the Poisson structure as \eqref{FT6}
\begin{equation}\label{BF5}
	\left[G_a,G_b\right]=f^c_{ab}G_c,\quad\left[G_a,G^b_{jk}\right]=-f^b_{ac}G^c_{jk},\quad\left[G^a_{jk},G^b_{lm}\right]=0,
\end{equation}
i.e., the submanifold defined by \eqref{BF3} is coisotropic within the same Poisson structure. It is immediately apparent that the first subset of constraints in \eqref{BF3}, which coincides with the first-class constraint set \eqref{YM7} of the Yang--Mills model, is irreducible. By contrast, the second subset coincides with the first-class constraint set \eqref{FT7} of the Freedman--Townsend model and therefore exhibits the same off-shell reducibility of order $(D-3)$, described by \eqref{FT17}--\eqref{FT20}.

Taking into account the preceding results concerning the phase space structure and the Hamiltonian gauge algebra, a straightforward counting argument shows that the model under consideration carries {\em no physical degrees of freedom}.  

At this stage, building on the preceding results, one can complete the gauge structure of the Lagrangian action associated with \eqref{BF1},
\begin{equation}\label{BF6}
	S^{\rm BF}:=\int{\rm d}t L^{\rm BF}(t).
\end{equation}
The construction starts from Dirac's conjecture concerning the role of first-class constraints as generators of gauge transformations and yields the following transformations of the phase space variable
\begin{align}
	\delta_\epsilon A^a_j&:=\left[A^a_j,G_m\epsilon^m+G^m_{kl}\epsilon^{kl}_m\right]=-(D_j)^a_{\hphantom{a}b}\epsilon^b,\label{BF7a}\\
	\delta_\epsilon B_a^{0k}&:=\left[B_a^{0k},G_m\epsilon^m+G^m_{lp}\epsilon^{lp}_m\right]=f^c_{ab}B^{0k}_c\epsilon^b+(D_l)_a^{\hphantom{a}b}\epsilon^{lk}_b.\label{BF7b}
\end{align} 
The gauge structure is then completed by the transformations of the Lagrange multipliers, here represented by the fields $A^a_0$ and $B^{jk}_a$. These transformations are determined by requiring the invariance of the action \eqref{BF6} under \eqref{BF7a}--\eqref{BF7b} and are given by
\begin{align}
	\delta_\epsilon A^a_0&=-(\dot{\epsilon}^a-f^a_{bc}A^c_0\epsilon^b),\label{BF8a}\\
	\delta_\epsilon B^{jk}_a&=f^c_{ab}B^{jk}_c\epsilon^b-(\dot{\epsilon}^{jk}_a+f^c_{ab}A^b_0\epsilon^{jk}_c)+(D_l)_a^{\hphantom{a}b}\epsilon^{ljk}_b,\label{BF8b}.
\end{align}
Introducing the redefined gauge parameters
\[
\bar{\epsilon}^a:=-\epsilon^a
\]
and the covariant, totally antisymmetric gauge parameters
\[
\bar{\epsilon}^{0jk}:=-\epsilon^{jk}_a,\quad\bar{\epsilon}^{jkl}:=\epsilon^{jkl}_a,
\]
one may recast infinitesimal local transformations \eqref{BF7a}--\eqref{BF8b} in a manifestly covariant form
\begin{equation}\label{BF9}
\delta_{\bar{\epsilon}}A^a_\mu=(D_\mu)^a_{\hphantom{a}b}\bar{\epsilon}^b,\quad\delta_{\bar{\epsilon}}B_a^{\mu\nu}=-f^c_{ab}B^{\mu\nu}_c\bar{\epsilon}^b+(D_\lambda)_a^{\hphantom{a}b}\bar{\epsilon}_b^{\lambda\mu\nu}.
\end{equation}
By direct computation it can be shown that the generating set of gauge transformations \eqref{BF9} closes off-shell, with the first-order structure functions being precisely the structure constants of the underlying Lie algebra
\[
\left[\delta_{\bar{\epsilon}},\delta_{\bar{\eta}}\right]A^a_\mu=\delta_{\bar{\xi}}A^a_\mu,\quad\left[\delta_{\bar{\epsilon}},\delta_{\bar{\eta}}\right]B^{\mu\nu}_a=\delta_{\bar{\xi}}B^{\mu\nu}_a,
\]
with
\[
\bar{\xi}^a:=f^a_{bc}\bar{\epsilon}^b\bar{\eta}^c,\quad \bar{\xi}_a^{\lambda\mu\nu}:=f^c_{ab}(\bar{\epsilon}^b\bar{\eta}_c^{\lambda\mu\nu}-\bar{\eta}^b\bar{\epsilon}_c^{\lambda\mu\nu}).
\]

Finally, the generating set of gauge transformations \eqref{BF9} is on-shell $(D-3)$-stage reducible. The corresponding hierarchy of reducibility relations is given by \eqref{FT17} and \eqref{FT19}, while the associated reducibility functions are listed in \eqref{FT18} and \eqref{FT20}, respectively.
\section{Non-linear theories}
The theories discussed in the previous section possess gauge algebras that close off-shell, with at most reducible gauge generators. A natural continuation is provided by nonlinear gauge theories, whose gauge structure is substantially richer. In these models, the commutator of two gauge transformations generally closes only modulo the equations of motion, giving rise to an open gauge algebra with field-dependent structure functions. Such theories play a prominent role in modern mathematical physics, encompassing Poisson sigma models \cite{Ikeda,Schaller,CattaneoFelder}, Courant Sigma models \cite{Roytenberg}, and various higher gauge theories, while also providing the natural setting for the Batalin--Vilkovisky quantization formalism \cite{AKSZK}.

The aim of this section is to investigate the extent to which the Faddeev--Jackiw procedure can be applied to this broader class of gauge systems. Particular emphasis will be placed on the derivation of the underlying geometric structures and on the characterization of the open gauge algebra within the Faddeev--Jackiw framework.

The nonlinear gauge theory under consideration is described by the Lagrangian function
\begin{equation}\label{NL1}
	L^{\rm NL}(t):=\int{\rm d}^{D-1}{\bf x}\left(H^a_\mu\bar{D}^\mu\varphi_a+\tfrac{1}{2}B^{\mu\nu}_a\bar{F}^a_{\mu\nu}\right),
\end{equation}
where
\begin{equation}\label{NL2}
	\bar{D}_\mu\varphi_a:=\partial_\mu\varphi_a+W_{ab}A^b_\mu,\quad\bar{F}^a_{\mu\nu}:=\partial_{[\mu}A^a_{\nu]}+\partial^aW_{bc}A^b_\mu A^c_\nu.
\end{equation}
The functions $W_{ab}$ are the coefficients of Poisson $2$-vector on the manifold parametrized by the scalar fields $\{\varphi_a:a=1,\cdots,N\}$,
\[
{\rm W}:=\tfrac{1}{2}W_{ab}\partial^a\wedge\partial^b,\quad\partial^a:=\frac{\partial}{\partial\varphi_a}
\]
and therefore satisfy the Jacobi identity
\begin{equation}\label{NL3}
	W_{m[a}\partial^mW_{bc]}=0.
\end{equation}

To implement the Faddeev--Jackiw procedure, one first performs the standard decomposition of the spacetime indices into their temporal and spatial components. After a straightforward integration by parts, the Lagrangian functional \eqref{NL1} assumes the form
\begin{equation}\label{NL4}
		L^{\rm NL}(t):=\int{\rm d}^{D-1}{\bf x}\left(B^{0j}_a\dot{A}^a_j+H^a_0\dot{\varphi}_a+A^a_0G_a+B^{jk}_aG^a_{jk}+H^a_j\gamma^j_a\right),
\end{equation}
where one employed the notations
\begin{equation}\label{NL5}
	G_a:=(\bar{D}_j)_a^{\hphantom{a}b}B^{0j}_b-W_{ab}H^b_0,\quad G^a_{jk}:=\tfrac{1}{2}\bar{F}^a_{jk},\quad\gamma^j_a:=\bar{D}^j\varphi_a,
\end{equation}
together with the covariant derivatives associated with the nonlinear theory,
\begin{equation}
(\bar{D}_\mu)^a_{\hphantom{a}b}=\partial_\mu\delta^a_b-\partial^a W_{bc}A^c_\mu,\quad(\bar{D}_\mu)_a^{\hphantom{a}b}=\partial_\mu\delta_a^b+\partial^b W_{ac}A^c_\mu,\quad
\end{equation}
By using the Faddeev--Jackiw identifications, \eqref{NL4} displays the following results.

First, at each spatial point, the phase space is parametrized by the fields $(B^{0j}_a,A^a_j,H^a_0,\varphi_a)$ and is endowed with a non-degenerate Poisson $2$-vector. The corresponding nonvanishing fundamental Poisson brackets are
\begin{equation}\label{NL6}
	\left[B^{0j}_a,A^b_k\right]=-\delta^j_k\delta^b_a,\quad\left[H^a_0,\varphi_b\right]=-\delta^a_b.
\end{equation}
Meanwhile, the remaining fields, namely $A^a_0$, $B^{jk}_a$, and $H^a_j$ act as Lagrange multipliers enforcing the constraints
\begin{equation}\label{NL7}
	G_a\approx0,\quad G^a_{jk}\approx0,\quad\gamma^j_a\approx0.
\end{equation} 

Second, as in non-Abelian BF theory, the nonlinear model under consideration exhibits trivial Hamiltonian dynamics, since its canonical Hamiltonian vanishes identically
\begin{equation}\label{NL8}
	H^{\rm NL}(t)=0.
\end{equation}

By direct computations, it can be shown that the surface \eqref{NL7} is coisotropic with respect to the symplectic structure associated with the Poisson one \eqref{NL6} as
\begin{align}
	\left[G_a,G_b\right]&=\partial^cW_{ab}G_c+\partial^{cd}W_{ab}B^{0j}_c\gamma_{dj},\label{NL9a}\\
	\left[G_a,G^b_{jk}\right]&=-\partial^bW_{ac}G^c_{jk}-\tfrac{1}{2}\partial^{bc}W_{ad}\gamma_{c[j}A^d_{k]},\label{NL9b}\\
	\left[G_a,\gamma^j_b\right]&=\partial^cW_{ab}\gamma^j_c,\label{NL9c}\\
	\left[G^a_{jk},G^b_{lm}\right]&=0,\quad\left[G^a_{jk},\gamma^l_b\right]=0=\left[\gamma^k_a,\gamma^l_b\right]\label{NL9d}.
\end{align}

The analysis of the constraints \eqref{NL7}, carried out using the identities
\begin{align}
	(\bar{D}_i)_a^{\hphantom{a}b}(\bar{D}_j)_b^{\hphantom{b}c}-(\bar{D}_j)_a^{\hphantom{a}b}(\bar{D}_i)_b^{\hphantom{b}c}&=\partial^{cd}W_{ab}\gamma_{d[i}A^b_{j]}+2\partial^c W_{ab}G^b_{ij}\nonumber\\
	&+W_{ab}\partial^{cd}W_{ef}A^e_i A^f_j,\label{NL10a}\\
	(\bar{D}_i)^a_{\hphantom{a}b}(\bar{D}_j)^b_{\hphantom{b}c}-(\bar{D}_j)^a_{\hphantom{a}b}(\bar{D}_i)^b_{\hphantom{b}c}&=-\partial^{ab}W_{cd}\gamma_{b[i}A^d_{j]}-2\partial^a W_{cd}G^d_{ij}\nonumber\\
	&-W_{cd}\partial^{ad}W_{ef}A^e_i A^f_j,\label{NL10b}
\end{align}
shows that they are on-shell $(D-2)$-stage reducible. The first-stage reducibility functions are
\begin{align}
	(\bar{Z}^a_{ijk})^{lm}_b&=\tfrac{1}{3!}(\bar{D}_{[i})^a_{\hphantom{a}b}\delta^l_j\delta^m_{k]},\quad& (\bar{Z}^a_{ijk})^b_l&=-\tfrac{1}{3!}\partial^{ab}W_{cd}A^c_{[i}A^d_j\eta_{k]l}\label{NL11a}\\
	(\bar{Z}_a^{ij})^{lm}_b&=-\tfrac{1}{2!}W_{ab}\eta^{i[l}\eta^{m]j},\quad& (\bar{Z}_a^{ij})^b_l&=\tfrac{1}{2!}(\bar{D}^{[i})_a^{\hphantom{a}b}\delta_l^{j]}\label{NL11b}
\end{align}
and satisfy the corresponding first-stage reducibility relations
\begin{equation}\label{NL12}
	(\bar{Z}^a_{ijk})^{lm}_b G^b_{lm}+(\bar{Z}^a_{ijk})^b_l\gamma^l_b=0,\quad
	(\bar{Z}_a^{ij})^{lm}_bG^b_{lm}+(\bar{Z}_a^{ij})^b_l\gamma^l_b=0.
\end{equation}
For each 
\[
4\leq p\leq D-1,
\]
the reducibility functions at stage  $L=p-2$ take the form 
\begin{align}
	(\bar{Z}^a_{i_1\cdots i_p})^{j_1\cdots j_{p-1}}_b&=\tfrac{1}{p!}(D_{[i_1})^a_{\hphantom{a}b}\delta^{j_1}_{i_2}\cdots\delta^{j_{p-1}}_{i_p]},\label{NL13a}\\
	(\bar{Z}^a_{i_1\cdots i_p})^{bj_1\cdots j_{p-2}}&=\tfrac{(-)^p}{p!}\partial^{ab}W_{cd}A^c_{[i_1}A^d_{i_2}\delta^{j_1}_{i_3}\cdots\delta^{j_{p-2}}_{i_p]},\label{NL13b}\\
	(\bar{Z}_a^{i_1\cdots i_{p-1}})_{bj_1\cdots j_{p-1}}&=\tfrac{(-)^p}{(p-1)!}W_{ab}\delta_{j_1}^{[i_1}\cdots\delta_{j_{p-1}}^{i_{p-1}]},\label{NL13c}\\
	(\bar{Z}_a^{i_1\cdots i_{p-1}})^b_{j_1\cdots j_{p-2}}&=\tfrac{1}{(p-1)!}(D^{[i_1})_a^{\hphantom{a}b}\delta_{j_1}^{i_2}\cdots\delta_{j_{p-2}}^{i_p]}.\label{NL13d}
\end{align}
The reducibility functions at the highest stage, $L=D-2$, are obtained from \eqref{NL13c} and \eqref{NL13d} by setting $p=D$.

For the same range of $p$, the reducibility relations at stage $L=p-2$ read  
\begin{align}
	&(\bar{Z}^a_{i_1\cdots i_{p+1}})^b_{k_1\cdots k_{p-1}}(\bar{Z}_b^{k_1\cdots k_{p-1}})_c^{j_1\cdots j_{p-1}}+(\bar{Z}^a_{i_1\cdots i_{p+1}})_b^{k_1\cdots k_p}(\bar{Z}^b_{k_1\cdots k_p})_c^{j_1\cdots j_{p-1}}\nonumber\\
	&=-\tfrac{1}{(p+1)!}\left(\partial^{ab}W_{cd}\gamma_{b[i_1}A^d_{i_2}\delta^{j_1}_{i_3}\cdots\delta^{j_{p-1}}_{i_{p+1}]}+2\partial^aW_{cd}G^d_{[i_1i_2}\delta^{j_1}_{i_3}\cdots\delta^{j_{p-1}}_{i_{p+1}]}\right),\label{NL14a}\\
	&(\bar{Z}^a_{i_1\cdots i_{p+1}})^b_{k_1\cdots k_{p-1}}(\bar{Z}_b^{k_1\cdots k_{p-1}})^c_{j_1\cdots j_{p-2}}+(\bar{Z}^a_{i_1\cdots i_{p+1}})_b^{k_1\cdots k_p}(\bar{Z}^b_{k_1\cdots k_p})^c_{j_1\cdots j_{p-2}}\nonumber\\
	&=\tfrac{(-)^p}{(p+1)!}\left(\partial^{acd}W_{ef}\gamma_{d[i_1}A^e_{i_2}A^f_{i_3}\delta^{j_1}_{i_4}\cdots\delta^{j_{p-2}}_{i_{p+1}]}+2\partial^{ac}W_{ef}G^e_{[i_1i_2}A^f_{i_3}\delta^{j_1}_{i_4}\cdots\delta^{j_{p-2}}_{i_{p+1}]}\right)\label{NL14b}\\
	&(\bar{Z}_a^{i_1\cdots i_p})^b_{k_1\cdots k_{p-1}}(\bar{Z}_b^{k_1\cdots k_{p-1}})_{cj_1\cdots j_{p-1}}+(\bar{Z}_a^{i_1\cdots i_p})_b^{k_1\cdots k_p}(\bar{Z}^b_{k_1\cdots k_p})_{cj_1\cdots j_{p-1}}\nonumber\\
	&=\tfrac{(-)^p}{p!}\partial^dW_{ac}\gamma_d^{[i_1}\delta_{j_1}^{i_2}\cdots\delta_{j_{p-1}}^{i_p]},\label{NL14c}\\
	&(\bar{Z}_a^{i_1\cdots i_p})^b_{k_1\cdots k_{p-1}}(\bar{Z}_b^{k_1\cdots k_{p-1}})^c_{j_1\cdots j_{p-2}}+(\bar{Z}_a^{i_1\cdots i_p})_b^{k_1\cdots k_p}(\bar{Z}^b_{k_1\cdots k_p})^c_{j_1\cdots j_{p-2}}\nonumber\\
	&=\tfrac{1}{p!}\partial^{bc}W_{ad}\gamma_c^{[i_1}A^{di_2}\delta_{j_1}^{i_3}\cdots\delta_{j_{p-2}}^{i_p]}+\tfrac{2}{p!}\partial^cW_{ad}G^{d[i_1i_2}\delta_{j_1}^{i_3}\cdots\delta_{j_{p-2}}^{i_p]}.\label{NL14d}	
\end{align}
Finally, the highest-stage reducibility relations, corresponding to $L=D-2$, are obtained from \eqref{NL14c} and \eqref{NL14d} by setting $p=D$.

In light of the preceding results concerning the phase space structure and the Hamiltonian gauge algebra, a straightforward counting argument shows that the model possesses \emph{no local physical degrees of freedom}, thereby confirming the topological character of the nonlinear theory under consideration.

At this stage, the Hamiltonian gauge structure can be used to reconstruct its Lagrangian counterpart. Accordingly, invoking the Dirac conjecture, one postulates the following infinitesimal gauge transformations of the phase space variables
\begin{align}
{\delta}_\epsilon B^{0j}_a&=\partial^cW_{ab}\epsilon^bB^{0j}_c+(D_k)_a^{\hphantom{a}b}\epsilon^{kj}_b+W_{ab}\xi^{bj},\label{NL15a}\\
		{\delta}_\epsilon A^a_j&=-(D_j)^a_{\hphantom{a}b}\epsilon^b,\label{NL15b}\\
		{\delta}_\epsilon H^a_0&=(\bar{D}^j)^a_{\hphantom{a}b}\xi^b_j+\partial^aW_{bc}H^c_0\epsilon^b-\partial^{ab}W_{cd}(B^{0k}_b\epsilon^c+\tfrac{1}{2}A^c_j\epsilon^{jk}_b)A^d_k,\label{NL15c}\\
		{\delta}_\epsilon \varphi_a&=W_{ab}\epsilon^b.\label{NL15d}
\end{align}   
These transformations are supplemented by those of the Lagrange multipliers,
\begin{align}
{\delta}_\epsilon A^a_0&=-(\bar{D}_0)^a_{\hphantom{a}b}\epsilon^b,\label{NL16a}\\
{\delta}_\epsilon B^{jk}_a&=-(\bar{D}_0)_a^{\hphantom{a}b}\epsilon^{jk}_b+(\bar{D}_l)_a^{\hphantom{a}b}\epsilon^{ljk}_b-W_{ab}\xi^{bjk}+\partial^cW_{ab}B^{jk}_c\epsilon^b,\label{NL16b}\\
{\delta}_\epsilon H^a_j&=-(\bar{D}^0)^a_{\hphantom{a}b}\xi_j^b+(\bar{D}^k)^a_{\hphantom{a}b}\xi_{kj}^b+\tfrac{1}{2}\partial^{ab}W_{cd}A^c_kA^d_l\epsilon_{bklj}+\partial^aW_{bc}H^b_j\epsilon^c\nonumber\\
&+\partial^{ab}W_{cd}A^{d\mu}B_{b\mu j}\epsilon^c+\partial^{ab}W_{cd}A^{dk}A^{c0}\epsilon_{bjk}.\label{NL16c}
\end{align}
The transformations \eqref{NL16a}--\eqref{NL16c} are determined by requiring the gauge invariance of the action
\begin{equation}\label{NL17}
	S^{\rm NL}:=\int{\rm d}t L^{\rm NL}(t)
\end{equation}
while taking into account the first-stage reducibility relations \eqref{NL12} satisfied by the first-class constraints \eqref{NL7}.

The additional gauge parameters $\epsilon^{jkl}_a$ and $\xi^a_{jk}$,  which are associated with the first-stage reducibility relations \eqref{NL12}, are understood to be completely antisymmetric in their respective spatial indices.

Introducing the Lorentz-covariant gauge parameters through the identifications
\[
\bar{\epsilon}^a:=-\epsilon^a,\quad \bar{\xi}^a_{\mu\nu}:=(-\xi^a_j,\xi^a_{jk}),\quad\bar{\epsilon}^{\mu\nu\lambda}_a:=(-\epsilon^{jk}_a,\epsilon^{jkl}_a), 
\]
where $\bar{\xi}^a_{\mu\nu}$ and $\bar{\epsilon}^{\mu\nu\lambda}_a$ are completely antisymmetric in their spacetime indices, the generating set of gauge transformations \eqref{NL15a}--\eqref{NL16c} can be recast in the manifestly covariant form
\begin{align}
	\delta_{\bar{\epsilon}}A^a_\mu&=(\bar{D}_\mu)^a_{\hphantom{a}b}\bar{\epsilon}^b,\label{NL18a}\\
	\delta_{\bar{\epsilon}}B_a^{\mu\nu}&=(\bar{D}_\lambda)_a^{\hphantom{a}b}\bar{\epsilon}^{\mu\nu\lambda}_b-W_{ab}\bar{\xi}^{b\mu\nu}-\partial^cW_{ab}B^{\mu\nu}_c\bar{\epsilon}^b,\label{NL18b}\\
	\delta_{\bar{\epsilon}}H^a_\mu&=(\bar{D}^\lambda)^a_{\hphantom{a}b}\bar{\xi}^b_{\lambda\mu}-\partial^aW_{bc}H^c_\mu\bar{\epsilon}^b+\partial^{ab}W_{cd}A^{c\lambda}(B_{b\lambda\mu}\bar{\epsilon}^d+\tfrac{1}{2}A^{d\rho}\bar{\epsilon}_{b\lambda\rho\mu}),\label{NL18c}\\
	\delta_{\bar{\epsilon}}\varphi_a&=-W_{ab}\bar{\epsilon}^b.\label{NL18d}
\end{align}

 By direct computations it can be shown that the generating set of gauge transformations \eqref{NL18a}--\eqref{NL18d} is highly non-linear. First, it closes on-shell, on the field equations corresponding to \eqref{NL17}, as
 \begin{align}
 	\left[\delta_{\bar{\epsilon}},\delta_{\bar{\epsilon}^\prime}\right]A^a_\mu&=\delta_{\bar{\epsilon}^{\prime\prime}}A^a_\mu-\partial^{ae}B_{bc}\frac{\delta S^{\rm NL}}{\delta H^{e\mu}}\bar{\epsilon}^b\bar{\epsilon}^{\prime c},\label{NL19a}\\
 	\left[\delta_{\bar{\epsilon}},\delta_{\bar{\epsilon}^\prime}\right]B_a^{\mu\nu}&=\delta_{\bar{\epsilon}^{\prime\prime}}B_a^{\mu\nu}-\partial^{bd}W_{ac}\frac{\delta S^{\rm NL}}{\delta H^{d\lambda}}(\bar{\epsilon}^c\bar{\epsilon}^{\prime\lambda\mu\nu}_b-\bar{\epsilon}^{\prime c}\bar{\epsilon}^{\lambda\mu\nu}_b),\label{NL19b}\\
 	\left[\delta_{\bar{\epsilon}},\delta_{\bar{\epsilon}^\prime}\right]H^a_\mu&=\delta_{\bar{\epsilon}^{\prime\prime}}H^a_\mu+\partial^{ab}W_{cd}\left(\frac{\delta S^{\rm NL}}{\delta A^{b\mu}}\bar{\epsilon}^c\bar{\epsilon}^{\prime d}-\frac{\delta S^{\rm NL}}{\delta B_{d\lambda\rho}}(\bar{\epsilon}^c\bar{\epsilon}^{\prime\lambda\mu\nu}_b-\bar{\epsilon}^{\prime c}\bar{\epsilon}^{\lambda\mu\nu}_b)\right)\nonumber\\
 	&-\frac{\delta S^{\rm NL}}{\delta H^e_\lambda}\left(\partial^{ae}W_{bc}(\bar{\epsilon}^b\bar{\xi}^{\prime c}_{\lambda\mu}-\bar{\epsilon}^{\prime b}\bar{\xi}_{\lambda\mu}^c)+\partial^{abe}W_{cd}B_{b\lambda\mu}\bar{\epsilon}^c\bar{\epsilon}^{\prime d}\right.\nonumber\\
 	&\left.\partial^{abe}W_{cd}A^{d\rho}(\bar{\epsilon}^c\bar{\epsilon}^{\prime}_{b\lambda\mu\nu}-\bar{\epsilon}^{\prime c}\bar{\epsilon}_{b\lambda\mu\nu})\right),\label{NL19c}\\
 	\left[\delta_{\bar{\epsilon}},\delta_{\bar{\epsilon}^\prime}\right]\varphi_a&=\delta_{\bar{\epsilon}^{\prime\prime}}\varphi_a,\label{NL19d}
 \end{align}
 where the notations have been considered
 \begin{align}
 	\bar{\epsilon}^{\prime\prime a}&:=\partial^aW_{bc}\bar{\epsilon}^b\bar{\epsilon}^{\prime c},\label{NL20a}\\
 	\bar{\xi}^{\prime\prime a}_{\mu\nu}&:=\partial^aW_{bc}(\bar{\epsilon}^b\bar{\xi}^{\prime c}_{\lambda\mu}-\bar{\epsilon}^{\prime b}\bar{\xi}_{\lambda\mu}^c)+\partial^{ab}W_{cd}\left(B_{b\mu\nu}\bar{\epsilon}^c\bar{\epsilon}^{\prime d}\right.\nonumber\\
 	&\left.-A^{d\lambda}(\bar{\epsilon}^c\bar{\epsilon}^{\prime}_{b\lambda\mu\nu}-\bar{\epsilon}^{\prime c}\bar{\epsilon}_{b\lambda\mu\nu})\right),\label{NL20b}\\
 	\bar{\epsilon}^{\prime\prime\lambda\mu\nu}_a&:=\partial^cW_{ab}(\bar{\epsilon}^b\bar{\epsilon}^{\prime\lambda\mu\nu}_c-\bar{\epsilon}^{\prime b}\bar{\epsilon}^{\lambda\mu\nu}_c).\label{NL20c}
 \end{align}
 
 Finally, the on-shell reducibility of the constraint set \eqref{NL7}, extending through stage $L=D-2$ and encoded in \eqref{NL13a}--\eqref{NL14d}, is inherited by the generating set of gauge transformations \eqref{NL18a}--\eqref{NL18d}. Indeed, in the standard DeWitt condensed notation, the generators associated with the aforementioned infinitesimal gauge transformations take the explicit form
 \begin{align}
 	(\bar{Z}^a_\mu)_b&=(\bar{D}_\mu)^a_{\hphantom{a}b}, &(\bar{Z}^{\mu\nu}_a)^b_{\alpha\beta\gamma}&=\tfrac{1}{3!}(\bar{D}_{[\alpha})_a^{\hphantom{a}b}\delta^\mu_\beta\delta^\nu_{\gamma]},\label{NL21a}\\
 	(\bar{Z}^{\mu\nu}_a)^{\alpha\beta}_b&=-\tfrac{1}{2}W_{ab}\eta^{\mu[\alpha}\eta^{\nu]\beta},&	(\bar{Z}^{\mu\nu}_a)_b&=-\partial^cW_{ab}B^{\mu\nu}_c,\label{NL21b}\\	(\bar{Z}_a)_b&=-W_{ab},&(\bar{Z}^{\prime a}_\mu)^{\alpha\beta}_b&=\tfrac{1}{2}(\bar{D}^{[\alpha})^a_{\hphantom{a}b}\delta^{\beta]}_\mu,\label{NL21c}\\
 	(\bar{Z}^{\prime a}_\mu)^b_{\alpha\beta\gamma}&=\tfrac{1}{3!}\partial^{ab}W_{cd}A^c_{[\alpha}A^d_\beta\eta_{\gamma]\mu},&(\bar{Z}^{\prime a}_\mu)_b&=\partial^aW_{cb}H^c_\mu+\partial^{ad}W_{cb}A^{c\lambda}B_{d\lambda\mu}.\label{NL21d}
 \end{align}
 Here, the primed gauge generators are those associated with the $1$-forms $H^a_\mu$.
 
Taking into account the gauge generators \eqref{NL21a}--\eqref{NL21d}, a direct computation based on the identities
 \begin{align}
 	(\bar{D}_\mu)_a^{\hphantom{a}b}(\bar{D}_\nu)_b^{\hphantom{b}c}-(\bar{D}_\nu)_a^{\hphantom{a}b}(\bar{D}_\mu)_b^{\hphantom{b}c}&=\partial^{cd}W_{ab}\frac{\delta S^{\rm NL}}{\delta H^{d[\mu}}A^b_{\nu]}+2\partial^cW_{ab}\frac{\delta S^{\rm NL}}{\delta B_b^{\mu\nu}}\nonumber\\
 	&+W_{ab}\partial^{cd}W_{ef}A^e_\mu A^f_\nu,\label{NL22a}\\
 	(\bar{D}_i)^a_{\hphantom{a}b}(\bar{D}_j)^b_{\hphantom{b}c}-(\bar{D}_j)^a_{\hphantom{a}b}(\bar{D}_i)^b_{\hphantom{b}c}&=-\partial^{ab}W_{cd}\frac{\delta S^{\rm NL}}{\delta H^{b[\mu}}A^d_{\nu]}-2\partial^a W_{cd}\frac{\delta S^{\rm NL}}{\delta B_d^{\mu\nu}}\nonumber\\
 	&-W_{cd}\partial^{ad}W_{ef}A^e_\mu A^f_\nu,\label{NL22b}
 \end{align}
shows that the reducibility functions at the successive stages can be written as
 \begin{align}
 (\bar{Z}_a^{\mu_1\cdots\mu_p})_{\alpha_1\cdots\alpha_{p+1}}^b&:=\tfrac{1}{(p+1)!}(\bar{D}_{[\alpha_1})_a^{\hphantom{a}b}\delta^{\mu_1}_{\alpha_2}\cdots\delta^{\mu_p}_{\alpha_{p+1}]},\quad p\leq D-1\label{NL23a}\\
 (\bar{Z}_a^{\mu_1\cdots\mu_p})_{b\alpha_1\cdots\alpha_p}&:=-\tfrac{(-)^p}{p!}W_{ab}\delta^{\mu_1}_{[\alpha_1}\cdots\delta^{\mu_p}_{\alpha_p]},\quad p\leq D\label{NL23b}\\
 (\bar{Z}^a_{\mu_1\cdots\mu_q})^{\alpha_1\cdots\alpha_{q+1}}_b&:=\tfrac{1}{(q+1)!}(\bar{D}^{[\alpha_1})^a_{\hphantom{a}b}\delta_{\mu_1}^{\alpha_2}\cdots\delta_{\mu_q}^{\alpha_{q+1}]},\quad q\leq D-1\label{NL23c}\\
 (\bar{Z}^a_{\mu_1\cdots\mu_q})^{b\alpha_1\cdots\alpha_{q+2}}&:=-\tfrac{(-)^q}{(q+2)!}\partial^{ab}W_{cd}A^{c[\alpha_1}A^{d\alpha_2}\delta_{\mu_1}^{\alpha_3}\cdots\delta_{\mu_q}^{\alpha_{q+2}]}.\quad q\leq D-2\label{NL23d}
 \end{align}
An analogous direct calculation then yields the higher-stage reducibility identities
\begin{align}
	&(\bar{Z}_a^{\mu_1\cdots\mu_p})^b_{\alpha_1\cdots\alpha_{p+1}}(\bar{Z}_b^{\alpha_1\cdots\alpha_{p+1}})_{c\beta_1\cdots\beta_{p+1}}+(\bar{Z}_a^{\mu_1\cdots\mu_p})_b^{\alpha_1\cdots \alpha_p}(\bar{Z}^b_{\alpha_1\cdots\alpha_p})_{c\beta_1\cdots\beta_{p+1}}\nonumber\\
	&=\tfrac{(-)^p}{(p+1)!}\partial^dW_{ac}\frac{\delta S^{\rm NL}}{\delta H^{d[\beta_1}}\delta_{\beta_2}^{\mu_1}\cdots\delta_{\beta_{p+1}]}^{\mu_p},\label{NL24a}\\
	&(\bar{Z}_a^{\mu_1\cdots\mu_p})_b^{\alpha_1\cdots\alpha_p}(\bar{Z}^b_{\alpha_1\cdots\alpha_p})^c_{\beta_1\cdots\beta_{p+2}}+(\bar{Z}_a^{\mu_1\cdots\mu_p})^b_{\alpha_1\cdots\alpha_{p+1}}(\bar{Z}_b^{\alpha_1\cdots\alpha_{p+1}})^c_{\beta_1\cdots\beta_{p+2}}\nonumber\\
	&=-\tfrac{1}{(p+2)!}\left(\partial^{dc}W_{ab}\frac{\delta S^{\rm NL}}{\delta H^{d[\beta_1}}A^b_{\beta_2}+2\partial^cW_{ad}\frac{\delta S^{\rm NL}}{\delta B_d^{[\beta_1\beta_2}}\right)\delta_{\beta_3}^{\mu_1}\cdots\delta_{\beta_{p+1}]}^{\mu_p},\label{NL24b}\\
	&(\bar{Z}^{a\mu_1\cdots\mu_p})^b_{\alpha_1\cdots\alpha_{p+2}}(\bar{Z}_b^{\alpha_1\cdots\alpha_{p+2}})^c_{\beta_1\cdots \beta_{p+3}}+(\bar{Z}^{a\mu_1\cdots\mu_p})_b^{\alpha_1\cdots\alpha_{p
	+1}}(\bar{Z}^b_{\alpha_1\cdots\alpha_{p+1}})^c_{\beta_1\cdots\beta_{p+3}}\nonumber\\
	&=\tfrac{(-)^p}{(p+3)!}\left(\partial^{abc}W_{de}\frac{\delta S^{\rm NL}}{\delta H^{b[\beta_1}}A^d_{\beta_2}+2\partial^{ac}W_{be}\frac{\delta S^{\rm NL}}{\delta B_b^{[\beta_1\beta_2}}\right)A^e_{\beta_3}\delta_{\beta_4}^{\mu_1}\cdots\delta_{\beta_{p+3}]}^{\mu_p},\label{NL24c}\\
	&(\bar{Z}^{a\mu_1\cdots\mu_p})^b_{\alpha_1\cdots\alpha_{p+2}}(\bar{Z}_b^{\alpha_1\cdots\alpha_{p+2}})_{c\beta_1\cdots\beta_{p+2}}+(\bar{Z}^{a\mu_1\cdots\mu_p})_b^{\alpha_1\cdots\alpha_{p+1}}(\bar{Z}^b_{\alpha_1\cdots\alpha_{p+1}})_{c\beta_1\cdots\beta_{p+2}}\nonumber\\
	&=\tfrac{1}{(p+2)!}\left(\partial^{ab}W_{ce}\frac{\delta S^{\rm NL}}{\delta H^{b[\beta_1}}A^e_{\beta_2}+2\partial^{a}W_{ce}\frac{\delta S^{\rm NL}}{\delta B_e^{[\beta_1\beta_2}}\right)\delta_{\beta_3}^{\mu_1}\cdots\delta_{\beta_{p+2}]}^{\mu_p}.\label{NL24d}
\end{align}
These identities complete the characterization of the Lagrangian gauge structure of the theory under consideration.
\section{Conclusions}

In this paper, the Faddeev--Jackiw procedure has been applied systematically to four representative non-Abelian gauge theories formulated on $D$-dimensional Minkowski spacetime: the Freedman--Townsend, Yang--Mills, and non-Abelian BF theories, together with a nonlinear topological model of BF type. The analysis demonstrates that, once the Lagrangian has been cast into first-order form, the Faddeev--Jackiw prescription provides a direct and economical means of identifying the phase space and its Poisson structure, the auxiliary variables and Lagrange multipliers, the coisotropic constraint submanifold, and the Hamiltonian generator of time evolution. It also yields the reducibility properties of the constraint set without requiring the complete sequence of steps characteristic of the standard Dirac--Bergmann algorithm.

For the Freedman--Townsend model, the procedure leads to an Abelian first-class constraint algebra that is on-shell $(D-3)$-stage reducible. The corresponding Hamiltonian is nonvanishing and generates the time evolution, while the counting of independent phase space variables and constraints yields $N$ local physical degrees of freedom, independently of the spacetime dimension. In the Yang--Mills case, the introduction of auxiliary fields provides an equivalent first-order formulation to which the Faddeev--Jackiw method applies directly. The resulting constraint algebra reproduces the Lie-algebraic commutation relations and is irreducible, leading to the expected $(D-2)N$ local physical degrees of freedom. The associated Lagrangian gauge algebra closes off-shell.

By contrast, the non-Abelian BF model possesses a vanishing canonical Hamiltonian and no local physical degrees of freedom, in agreement with its topological character. Its Hamiltonian constraint algebra is non-Abelian, while the corresponding generating set of Lagrangian gauge transformations closes off-shell. Nevertheless, this generating set is on-shell $(D-3)$-stage reducible. Thus, the off-shell closure of the gauge algebra does not preclude the on-shell character of its reducibility relations.

The nonlinear BF-type model demonstrates that the same procedure remains effective beyond gauge theories governed by Lie-algebraic structure constants. In this case, the Poisson tensor defined on the target space gives rise to field-dependent structure functions. The resulting coisotropic constraint algebra is genuinely nonlinear, the canonical Hamiltonian vanishes identically, and the theory carries no local physical degrees of freedom. At the Lagrangian level, the gauge algebra is open: the commutator of two gauge transformations closes only modulo the Euler--Lagrange equations. Furthermore, the generating set of gauge transformations is on-shell $(D-2)$-stage reducible.

For all four models, the Hamiltonian information furnished by the Faddeev--Jackiw analysis was supplemented by the Dirac conjecture in order to reconstruct the corresponding Lagrangian gauge transformations. Requiring invariance of the action then determined the transformations of the Lagrange multipliers and enabled the resulting gauge symmetries to be recast in manifestly covariant form. This construction also showed how the reducibility properties identified at the Hamiltonian level are inherited by the Lagrangian gauge generators.

Taken together, these results establish the Faddeev--Jackiw approach as a unified geometric framework for extracting both Hamiltonian and Lagrangian gauge structures from a broad class of non-Abelian theories. In particular, the method applies not only to conventional theories with closed Lie-algebraic gauge algebras, but also to topological and nonlinear models characterized by higher-stage reducibility and open, field-dependent gauge algebras. Possible extensions of the present analysis could include theories on curved backgrounds.
\section*{ORCID}

\noindent Eugen-Mihaita CIOROIANU - \url{https://orcid.org/0000-0002-7030-9770}

\end{document}